\documentclass[journal=jacsat,manuscript=article]{achemso}

\usepackage[version=3]{mhchem} 
\usepackage{xcolor}
\mciteErrorOnUnknownfalse

\author{A N M Nafiul Islam}
\affiliation[1]
{School of Electrical Engineering \& Computer Science, The Pennsylvania State University, University Park, 16802, PA, USA}
\author{Xuezhong Niu}
\author{Jiahui Duan}
\author{Shubham Kumar}
\author{Kai Ni}
\affiliation[2]
{Department of Electrical Engineering, University of Notre Dame, Notre Dame, 46556, IN, USA}
\author{Abhronil Sengupta}
\affiliation[1]
{School of Electrical Engineering \& Computer Science, The Pennsylvania State University, University Park, 16802, PA, USA}
\email{sengupta@psu.edu}
\title[An \textsf{achemso} demo]
  {Dendritic Computing with Multi-Gate Ferroelectric Field-Effect Transistors}

\abbreviations{IR,NMR,UV}

\begin{document}







\begin{abstract}
  Although inspired by neuronal systems in the brain, artificial neural networks generally employ point-neurons, which offer far less computational complexity than their biological counterparts. Neurons have dendritic arbors that connect to different sets of synapses and offer local non-linear accumulation -- playing a pivotal role in processing and learning. Inspired by this, we propose a novel neuron design based on a multi-gate ferroelectric field-effect transistor that mimics dendrites. It leverages ferroelectric nonlinearity for local computations within dendritic branches, while utilizing the transistor action to generate the neuronal output. The branched architecture enables smaller crossbar arrays in hardware integration, improving efficiency. Using an experimentally calibrated device-circuit-algorithm co-simulation framework, we demonstrate that networks incorporating our dendritic neurons achieve superior performance compared to much larger networks without dendrites ($\sim17\times$ fewer trainable weight parameters). These findings suggest dendritic hardware can significantly improve computational efficiency and learning capacity of neuromorphic systems optimized for edge applications.
\end{abstract}


\section{}
Artificial Intelligence (AI) has become ubiquitous and has revolutionized countless fields. The artificial neural networks (ANNs) at the heart of deep learning algorithms that power these AI systems generally employ a simplified neuron model known as the point neuron\cite{mcculloch1943logical}, where a linear summation of all synaptic outputs is performed, followed by a single lumped nonlinear activation function, such as a rectified linear unit, ReLU. This abstraction of the neuron ignoring spatial or structural nuances is vastly different to that seen in nature. In fact, biological neurons exhibit far more complicated functionalities beyond this basic somatic behavior. 

\begin{figure}
\centering
\includegraphics[width=0.85\textwidth]{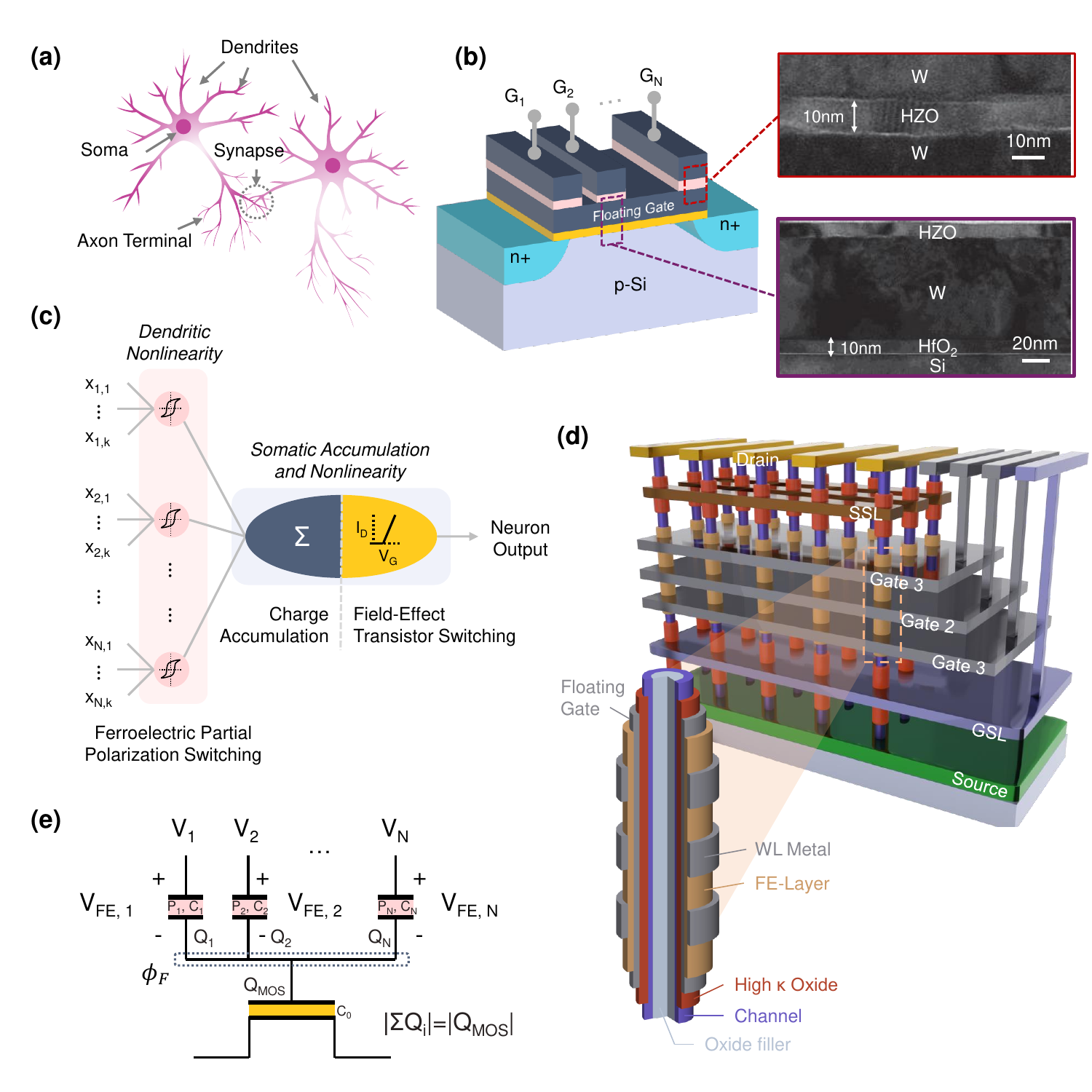}
\caption{\textbf{Overview of Dendritic Computing with Multi-gate FeFET.} (a) Signal from one neuron's axon terminals travel to the next neuron's dendrites through the synaptic cleft. It reaches the neuron's cell body or soma and is then propagated forward. (b) An n-gate FeFET structure along with the Transmission Electron Microscopy (TEM) cross-sectional images of the material stack. (c) The dendritic nonlinearity is mapped to the partial polarization switching of the ferroelectric layer, while the somatic accumulation occurs through charge accretion at the shared floating gate. The somatic nonlinearity is mapped to the FET switching characteristics. (d) The scalability and back-end-of-line compatibility of the ferroelectric layers enables us to integrate the multi-gate FeFET device vertically in 3D architectures to ensure high density. The devices are interfaced with ground select transistors/line (GSL) and string select transistors/line (SSL) for proper operation. (e) The multi-gate device symbol. Its structure can be abstracted as multiple ferroelectric capacitors connected in parallel with a FET in series connection below. The partial polarization switching characteristics induces a non-linear component to the floating gate voltage ($\phi_F$), which then modifies the channel characteristics to obtain the modulated drain current, the final neuronal output.}\label{device}
\end{figure}

A key feature of real neurons is the presence of dendrites -- branch-like structures that receive and locally process synaptic inputs before contributing to the neuron’s overall output (Fig.\ref{device}(a)). These dendrites are not merely passive conduits; rather, they actively shape input signals through nonlinear, location-specific interactions\cite{stuart2016dendrites,poirazi2001impact,koch1983nonlinear}. Numerous studies have shown that the local plasticity offered by dendrites greatly improves the information processing ability of biological networks and may underlie the brain's superior learning ability\cite{poirazi2003pyramidal,wu2009capacity,tzilivaki2019challenging}. Recent computational efforts in ANNs have shown that these benefits in the biological realm do indeed translate to quantifiable improvements in learning accuracy, generalization, and adaptability across real-world tasks\cite{hussain2014improved,wu2018improved,jones2021might,wang2022dendritic,zhang2024dendritic,chavlis2025dendrites,liu2024dendritic}.

Parallel to algorithmic developments, efforts in developing neuromorphic computing hardware have tried to create devices that can mimic the behavior of the brain primitives. Like ANNs, however, they also primarily focus on synaptic and somatic neuronal devices -- largely overlooking dendrites. Some prior works have attempted to model dendrites using discrete devices\cite{li2020power,liu2024graphene,d2024denram,boahen2022dendrocentric,chen2023multi}, circuit-level demonstrations\cite{edwards2024neural,ramakrishnan2013neuron}, or simplified architectural modifications\cite{hussain2014improved,roy2016online}. While these approaches have provided useful insights, they often remain constrained to shallow networks or are limited to processing temporal patterns. They offer no obvious path forward for scaling to deeper neural networks that pioneer the current AI advancements. Furthermore, these methods generally require the inclusion of significant additional peripheral circuitry – sacrificing energy and area efficiency. Thus, the hardware advantages of incorporating dendrites into neural network hardware remain unclear. 

In this work, we attempt to answer these questions and propose a singular device that can be employed as a neuron with integrated dendritic branches, operating in the analog regime. Our device is designed to be seamlessly incorporated in crossbar array architectures, which are widely used in neuromorphic systems for their parallelism and density – accelerating efficient computation in deep neural networks. The proposed neuron is based on a multi-gate Ferroelectric Field Effect Transistor (FeFET), shown in Fig.\ref{device}(b). The 1T-nC device structure consists of a metal-ferroelectric-metal-insulator-semiconductor (MFMIS) stack. The input gates, which represent the dendrite branches, have a thin film Hafnium Zirconium Oxide (HZO) ferroelectric layer with a floating metal gate underneath, shared by all the gates. Below the shared electrode is the thin dielectric oxide layer (HfO\textsubscript{2}) followed by the transistor channel, with its source and drain. 

The device architecture enables distinct functionalities at different regions. As shown in Fig.\ref{device}(c), the non-linear switching characteristic of the ferroelectric layer is utilized as the dendritic non-linearity, while the field effect switching of the channel underneath performs the somatic operation. When a drain voltage is applied, the drain current works as the output of the entire neuron. Channel conductivity is determined by the floating gate potential, which is dependent on the ferroelectric capacitors. The partial polarization switching of the domains in the ferroelectric material under various amplitudes of programming pulses leads to controlled non-linear charge accumulation on the floating gate, which then exerts its influence in the underlying channel and its conduction characteristics. This mechanism yields a fundamentally different functionality compared to the NAND structure or any previously discussed multi-gate FeFET\cite{chen2023multi,lee2022novel,deng2022compact}. However, like 3D-NAND, the structure can leverage the excellent scalability\cite{lee2021unveiling} and back-end-of-line (BEOL) compatibility\cite{aabrar2022beol} of the ferroelectric layers to enable 3D integrated architectures (Fig.\ref{device}(d)). Coupled with FeFET based synapses\cite{saha2021intrinsic,islam2023hybrid}, the multi-gate FeFET can realize an efficient, adaptive and scalable monolithic neuromorphic platform.   

Building on this device innovation, we leverage the intrinsic nonlinearity to design a dendritic deep neural network. We experimentally calibrate a device model with 3-gates and extend it for n-gates and use it to develop a device-circuit-algorithm co-simulation framework to assess the end-to-end impact of the dendritic computation on learning tasks. Our results clearly delineate the advantages afforded by the proposed device – both from a performance and hardware-efficiency perspective and paves a scalable path for next-generation neuromorphic systems that bring AI closer to the brain’s computing paradigm.



We start by analyzing the functionality of the proposed device. Fig.\ref{device}(e) shows different gate voltages being applied across the different gate terminals of the device. We denote the terminal voltage, the voltage and charge across the ferroelectric layer of the $i$-th gate as $V_i$, $V_{FE,i}$  and $Q_i$  respectively. $\phi_F$ and $Q_F$ are the floating gate voltage and charge respectively. $Q_{MOS}$ is the charge across the shared floating gate and the channel. Assuming no charge injection occurring during device operation\cite{shibata2002functional}, we can write the following equation by equating the charges:

\begin{equation}
Q_F = Q_{MOS} +\sum_{i=1}^N (-Q_i)
\end{equation}

Here, $N$ is the total number of gates. The charge across the ferroelectric layer is a summation of the polarization, $P_i$ and the product of the ferroelectric capacitance, $C_i$ and $V_{FE,i}$. Similarly, $Q_{MOS}$ can be written as a product of the floating gate voltage and the oxide capacitance, $C_0$. \textcolor{black}{Note, $C_0$ is not constant, but varies based on the operating regime of the underlying FET.} Assuming the initial charge at the floating gate to be zero (i.e., $Q_F=0$), we can now rewrite (1) as follows:

\begin{equation}
        C_0 \cdot \phi_F = \sum_{i=1}^N (P_i + C_i \cdot V_{FE,i})\\
        = \sum_{i=1}^N (P_i + C_i \cdot (V_i- \phi_F ))\\
        = \sum_{i=1}^N (P_i + C_i \cdot V_i) - \phi_F \sum_{i=1}^N C_i
\end{equation}

Solving for $\phi_F$, we get,

\begin{equation}
\phi_F = \frac{\sum_{i=1}^N C_i \cdot V_i}{\sum_{i=0}^N C_i} + \frac{\sum_{i=1}^N P_i}{\sum_{i=0}^N C_i}
\end{equation}

Here, $\phi_F$ consists of two terms: one depending on the capacitances while the other depends on the polarization of the ferroelectric layer. It is this latter term that is responsible for the non-linearity of dendrites. \textcolor{black}{Discussion on non-zero floating gate voltage is in SI.}

\textcolor{black}{Now, in order to calculate the polarization of the $i$-th gate for (3), we assume that the ferroelectric layer consists of multiple independent domains, $N_{dom}$ and can then employ a Monte Carlo algorithm based ferroelectric model\cite{saha2021intrinsic,deng2020comprehensive}}. Each domain in the ferroelectric layer can be polarized in either one of the two stable orientations, generally denoted by $(-1,1)$. Under a time-varying applied electric field, $E_{FE}(t)$, the switching probability, $p_k$, of the $k$-th domain with a certain time step, $\Delta t$, can be expressed by\cite{alessandri2019monte}, 

\begin{equation}
p_k (t_s < t + \Delta t | t_s > t) = 1-exp\left[h_k(t)^\beta-(h_k(t+\Delta t))^\beta\right]
\end{equation}

Here, $t_s$ is the switching time of the $k$-th domain considering the domain has not switched before $t$; $\beta$ is the shape parameter of the probability distribution of the activation fields, $E_{a,k}$; $h_k$ is the history parameter that tracks the change of the switching time constant, $\tau_k$, given by\cite{alessandri2019monte}:

\begin{equation}
h_k (t) = \int_{t_0}^t \frac{dt'}{\tau_k (E_{fe}(t'),E_{a,k})} 
\end{equation}

\begin{equation}
\tau_k (E_{fe}, E_{a,k}) = \tau_0  exp \left[ (\frac{E_{a,k}}{E_{fe}})^\alpha \right]
\end{equation}

The total polarization of each gate's ferroelectric layer, $P_{total,i}$, can be measured by summing up the states of all of the domains at any time, $t$ by:

\begin{equation}
P_{total,i} (t) = \frac{P_S}{M_i} \sum_{k=1}^{M_i} s_k (t)
\end{equation}

Here, $M_i$ is the total number of domains in the $i$-th gate's ferroelectric layer; $P_S$ is the saturated polarization of the ferroelectric and $s_k$ denotes the state of the domain (i.e., $-1$ or $+1$). \textcolor{black}{By combining the ferroelectric polarization model with the transistor charge-voltage equations (1-3), we can clearly see the dendritic non-linearity arises from the partial polarization switching in the ferroelectric. We can calculate the final resultant drain current, i.e.  the final output of our neuron, by self-consistently solving the transistor-charge equations.}

\begin{figure}
\centering
\includegraphics[width=\textwidth]{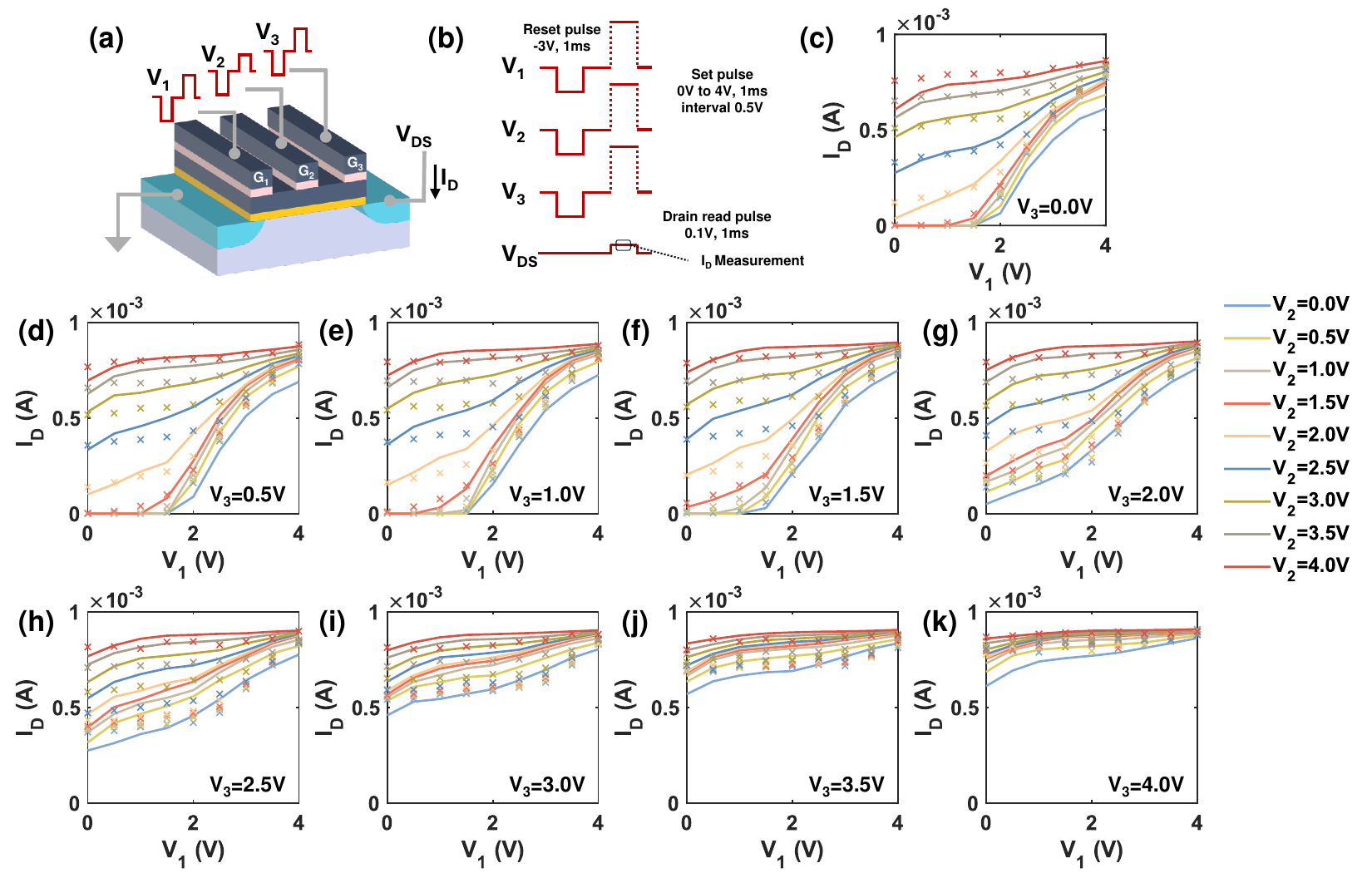}
\caption{\textbf{Characterization of Neuronal Behavior of the Multi-gate FeFET.} (a) Experimental setup for measuring the multi-gate FeFET characteristics (3-gates). (b) Each gate receives a reset pulse followed by a set/programming pulse, during which the drain current is measured for an applied drain voltage, $V_{DS}$. (c-k) Drain current $I_D$ with respect to applied set voltage to the first gate, $V_1$. For each panel, the different colored curves represent different set pulse values applied to the second gate, $V_2$. Likewise, the set voltage applied to the third terminal, $V_3$ remains constant for each panel and is noted at the bottom right of the panel. It should be noted that the gates are interchangeable, and the same results can be obtained by selecting any other relative combination. The calibrated model (solid lines) matches well with the experimental data (represented with $\times$’s).}\label{characterization}
\end{figure}


To validate and calibrate the discussed model, we fabricated and characterized a 3-gate FeFET device, as shown in Fig.\ref{characterization}(a). Details regarding device fabrication and the measurement setup can be found in the Supplementary Information (SI). We begin by validating the ferroelectric switching behavior of our ferroelectric capacitors (Supplementary Fig.\ref{S1}-\ref{S3}) and the FET switching behavior (Supplementary Fig.\ref{S4}). Next, we measure the device's response to programming voltages of varying amplitudes applied to each gate (Fig.\ref{characterization}(b)). Initially a reset pulse of $-3V$ for $1ms$ is applied to each of the gates to reset all the domains of the ferroelectric layer to the negative polarization state. We follow the reset pulse with a programming/set pulse voltage for $1ms$. The magnitude of these pulses varies from $0V$ to $4V$ with increments of $0.5V$ to all gates, i.e., we apply set pulses of all possible combinations on the three gates (total of $9\times9\times9=729$ individual combinations). The final drain current is measured by applying a $1ms$ pulse of $V_{DS}=0.1V$ to obtain the output of the neuron. Fig.\ref{characterization}(c-k) shows how drain current, $I_D$ changes with respect to one of the gate voltages ($V_1$) as the other two gate voltages ($V_2/V_3$) are varied from $0V$ to $4V$, with an interval of $0.5V$. \textcolor{black}{Now to fit our model with the experiment, we first tune our ferroelectric capacitor model to the ferroelectric switching characteristics (Supplementary Fig. S1 and S3). Next, we combine them with the FET parameters to fit the model results with outputs obtained under same operating conditions. The complete model parameters are listed in Supplementary Table \ref{dev_params}}. The corresponding floating gate voltages for the different drain currents are shown in Supplementary Fig.\ref{S5}.  

\begin{figure}
\centering
\includegraphics[width=0.8\textwidth]{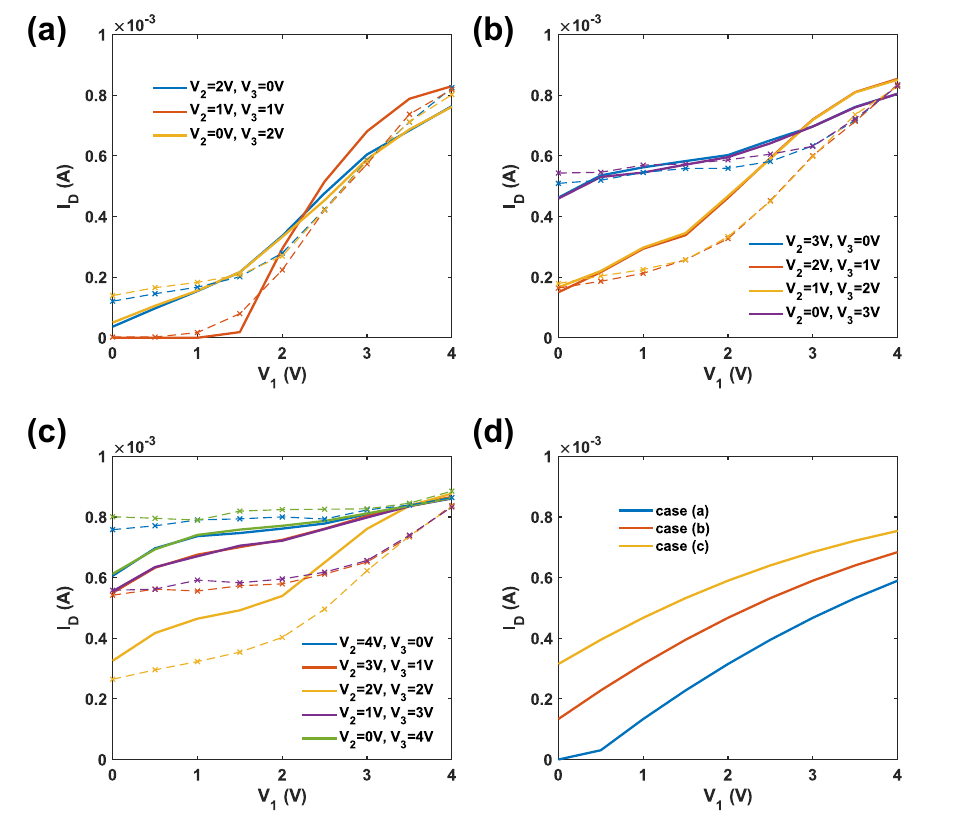}
\caption{\textbf{Impact of Ferroelectric Non-linear Switching.} Drain current, $I_D$ as one of the gate voltages ($V_1$) is swept from $0V$ to $4V$ for different amplitudes of $V_2$ and $V_3$ for our multi-gate dendritic device. \textcolor{black}{Solid and dashed lines represent simulated and experimental results respectively.} (a) $V_2=2V, V_3=0V$; $V_2=1V, V_3=1V$ and  $V_2=0V, V_3=2V$, (b) $V_2=3V, V_3=0V$; $V_2=2V, V_3=1V$; $V_2=1V, V_3=2V$ and $V_2=0V, V_3=3V$, (c) $V_2=4V, V_3=0V$; $V_2=3V, V_3=1V$; $V_2=2V, V_3=2V$; $V_2=1V, V_3=3V$ and $V_2=0V, V_3=4V$. Due to the nonlinear interactions, the $I_D$ curves vary for the different cases. Note, all the gates are interchangeable. (d) Drain current, $I_D$ vs. gate voltage $V_1$ with purely capacitive dendritic gates for the input voltage cases shown in (a) – (c). The inputs overlap as there is no non-linear component in the accumulation.}\label{impact}
\end{figure}

With the calibrated model, we can now compare the non-linear behavior of the device due to the partial polarization switching to a similar device with only capacitive coupling in its gates. To elucidate the importance of ferroelectric switching, we will look at how the device $I_D$ changes with respect to one of the gate voltages ($V_1$) for different values of voltage set pulses applied to the other gates ($V_2, V_3$). Fig.\ref{impact}(a—c) shows the $I_D$ vs $V_1$ plots for three different sets of inputs. While the linear sum of the inputs $V_2$ and $V_3$ are all same for the three plots, the drain current varies significantly. In Fig.\ref{impact}(a), we see that when $V_2$ and $V_3$ are small ($1V$), the device's threshold switching with respect to $V_1$ occurs at $\sim1.5V$. However, when either $V_2$ or $V_3$ has a larger value ($2V$), more ferroelectric domains are able to switch in those gates, and thus the threshold with respect to $V_1$ is effectively lowered -- resulting in non-zero drain current even at $V_1=0V$. Interestingly, in Fig.\ref{impact}(a), we see that although the device turns on later with respect to $V_1$ when $V_2=V_3=1V$, the drain current can actually be larger than the other two cases beyond $V_1>2V$. This can be explained by the fact that the larger gate voltage in $V_2/V_3$ makes a greater contribution to the shared floating gate voltage ($\phi_F$) and thus reduces the effective voltage across the ferroelectric layer of the first gate. This then results in fewer polarization switches and a lower drain current.
Similar discussions hold true for the cases shown in Fig.\ref{impact}(b) and (c) as well. Note, for our device all the gates are interchangeable as reflected by the different curves with same magnitudes but different order overlapping. In a purely capacitive setting, all the different voltages in the three different cases discussed above result in a simplistic summation of charges in the shared electrode (no non-linear polarization switching) and thus result in the same drain current characteristics. This can be simulated by turning off the polarization switching in our model by setting $P_S=0$ and the results are shown in Fig.\ref{impact}(d) – confirming our hypothesis. Further insights into the device behavior, drawn from varying various device and material parameters in the model, can be found in the SI and Supplementary Figs. S6--S9. 

\begin{figure}
\centering
\includegraphics[width=\textwidth]{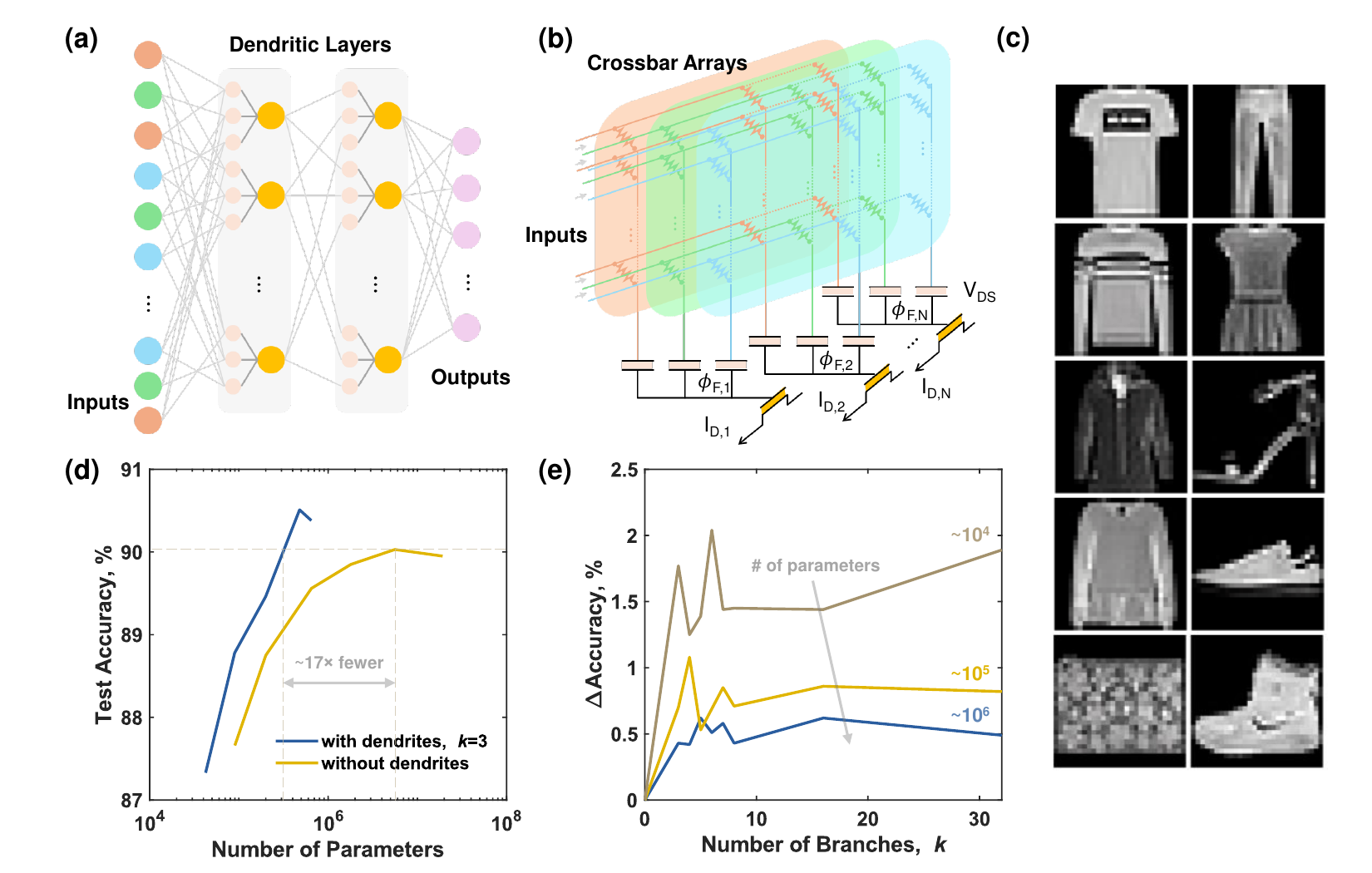}
\caption{\textbf{Network-level Performance evaluation.} (a) The proposed deep dendritic neural network architecture with $k=3$ dendritic branches. In each dendritic layer, each branch is connected to a subset of the previous layer. We have color-coded the input layer to show how they are connected to different branches. These dendritic layers can be stacked to create deep architectures. (b) The circuit architecture of the proposed dendritic layer. Each color, like (a), denotes a separate crossbar array that is interfaced with different gates of our device. For applied drain voltage, $V_{DS}$, the resultant drain currents are the outputs for the next layer. This architecture allows us to optimize our device based on the crossbar array size. \textcolor{black}{Additional current-to-voltage converters, not shown for sake of simplicity, will be needed to interface the gates to the crossbar columns.} (c) The Fashion-MNIST dataset, consisting of grayscale images assorted into 10 categories, was used for training and evaluating the network-level performance of networks with and without dendrites. (d) Test accuracy of networks with and without dendrites. Dendritic functionality is enabled by the multi-gate FeFET. We find that the network with dendrites is able to achieve equivalent performance to much larger (in terms of trainable weight parameters) networks without dendrites. (e) Performance improvements ($\Delta Accuracy$) on the test set, with respect to a network with equal trainable weight parameters but no dendrites. It is revealed that the efficacy of the dendrites increases as the total number of parameters decreases.}\label{network}
\end{figure}

Next, we incorporate this device model into a co-simulation framework to investigate the co-design possibilities and potential of the device in network-level learning scenarios.
The multi-gate device structure lends itself to being integrated directly into crossbar implementations of neural networks similar to prior neuronal device implementations. Fig.\ref{network}(a,b) shows the proposed network and its corresponding architectural circuit implementation. \textcolor{black}{The output current from each column of the crossbar are mapped and converted to corresponding voltages and are fed as inputs for the different gates of the device. This effectively decouples the crossbar and places no restrictions on its operation. Further details regarding the mapping can be found in the SI.}  

\textcolor{black}{Now, i}nstead of all the weights of a column being connected to a single neuronal device, each gate is connected to a subset of weights, i.e., if there are $d$ dendritic branches/input gates and $n$ weights per column of the crossbar, each gate is connected to $k=n/d$ connections randomly selected without replacement. Thus, each synaptic weight is only connected to a single input gate. Note, this only needs to be done once during network initialization and therefore does not incur any overhead in hardware implementation. Now, as each gate is connected to only $k$ weights, instead of $n$, this significantly reduces the required crossbar array size. In reality, crossbar size is dependent on various factors\cite{jain2018rx} and as they are scaled, non-idealities in the device-level (stochastic write operations, process variations) and circuit-level (driver/sensing resistance, sneak paths, interconnect parasitics, analog-to-digital/digital-to-analog converter nonlinearity) compound, leading to inefficiencies and erroneous operation. Thus, instead of a single large crossbar, it is generally broken into multiple crossbar arrays, whose outputs are later summed together. This summation operation is generally performed in the digital domain, thereby requiring costly analog-to-digital converters and storage of intermediate values – eschewing the efficiency gains of in-memory computation. In comparison, our device and algorithm framework allows us to avoid this by performing all computations in the analog domain, while preserving the benefits afforded by the use of smaller crossbar arrays. 

The inputs are applied concurrently across the gates. Mathematically, the floating gate voltage can be written as: 

\begin{equation}
\phi_F \propto g \left[ \sum_{l=1}^d \textbf{S}_l \odot \textbf{W} \odot \textbf{x} \right]     
\end{equation}

Here, $x$ is the incoming input to the crossbar, $W$ denotes the weight vector of the crossbar column, $S_l$ is a binary masking vector that indicates whether the weight is connected to the $l$-th branch and $g[\cdot]$ is the dendritic non-linearity offered by the ferroelectric. \textcolor{black}{Details regarding the abstraction of the dendritic and somatic non-linearities for training of the network is presented in SI and Supplementary Fig.\ref{S4} and \ref{S10}. The impact of the device characteristics on network-level inference tasks is evaluated next.}

We analyze the proposed network for learning tasks by combining our device model into a device-circuit-algorithm co-simulation framework. We train the networks on the Fashion-MNIST\cite{xiao2017fashion} dataset (Fig.\ref{network}(c)). The network consisted of $28\times28=784$ input neurons followed by $2$ hidden layers with each layer having $P$ neurons with $k$ dendritic branches for each neuron and an output layer with $10$ neurons. The neurons in the hidden layers utilize our developed device model. When $k$ is equal to zero, it represents a network with neurons with no dendritic branches. Details regarding the simulation can be found in the SI.

We start off by comparing the results of a network with dendrites ($k=3$) using the multi-gate FeFETs to a network which does not employ dendrites ($k=0$) – ideal software ReLU neurons. Note, both the types of networks employ the same architecture with equivalent weight parameters and only differ in their neuronal operation. As illustrated in Fig.\ref{network}(d), we find that as the number of trainable weight patterns are varied, the resultant test accuracy of networks without dendrites achieve a maximum accuracy of 90.03\% and decrease as the number of parameters is reduced. In contrast, the dendritic network is able to achieve better accuracies with networks with fewer number of parameters. In fact, we find that the maximum accuracy obtained by the networks without dendrites can be achieved by a dendritic network that uses approximately $17\times$ fewer parameters ($>1$ order of magnitude improvement). Further improvements are possible for dendritic networks by sampling the weights to the different branches using a variety of strategies, rather than random selection\cite{chavlis2025dendrites}. This result highlights that employing the multi-gate FeFETs can significantly cut down on required resources (chip area, power, etc.) for computing – making them ideal for applications related to edge intelligence. 

Next, we explored how network accuracy changes as the number of parameters and number of branches, $k$, are varied independently (Fig.\ref{network}(e)). We compare the difference of test accuracies with an equivalent network (same number of trainable parameters) with no dendrites ($k=0$) and find that the performance scales and saturates as the number of branches is increased. It is also observed that the improvement owing to dendritic branches are greater as the number of parameters are decreased, further elucidating the impact of the dendritic network. \textcolor{black}{To explain this apparent advantage of dendritic networks, we present a preliminarily exploration of how dendrites might impact network's complexity in SI and Supplementary Fig.\ref{S11}.}

It should be noted that device-level non-idealities (inter-gate parasitics, leakage, sneak paths, etc.) can become a factor as $k$ is increased, which can lead to errors in operation. We have not considered such effects for our first-order analysis in this work. \textcolor{black}{Similarly, endurance and fatigue of the ferroelectric capacitors also require careful consideration. While traditional memory applications require substantial retention ($>10$ years), for our case, switching endurance needs highest priority (for proper operation, each gate capacitor is reset before each set operation/forward pass). This can lead to significantly different material design considerations \cite{hur2020interplay,song2022positive}. While currently reported endurance of HZO capacitors ($\sim10^{11}$ cycles) are sufficient for most inference operations\cite{kozodaev2019mitigating,lyu2020high}, improvements stemming from these new considerations can further increase device longevity.}


In summary, we propose a multi-gate FeFET as a neuron with dendritic branches. We incorporate an experimentally calibrated device model of the neuron into a device-circuit-algorithm co-simulation framework and clearly highlight the advantage afforded by such a hardware implementation in comparison to networks without dendrites. This work in combination with prior efforts on ferroelectric synapses\cite{saha2021intrinsic,islam2023hybrid} can pave the path forward for a fully integrated learning framework for future resource constrained edge devices.

\begin{acknowledgement}

The authors would like to acknowledge GlobalFoundries Dresden Germany for providing FeFET testing devices.\\
\noindent
\textbf{Funding:} This material is based upon work supported in part by the U.S. National Science Foundation under award No. CNS \#2137259 - Center for Advanced Electronics through Machine Learning (CAEML) and its industry members, \#2318101 and \#2346953. The electrical characterization is supported by the U.S. Department of Energy, Office of Science, Office of Basic Energy Sciences Energy Frontier Research Centers program under Award Number DESC0021118.\\
\noindent
\textbf{Conflict of interest:} Authors declare that they have no competing interests.\\
\noindent
\textbf{Data and materials availability:} All data needed to evaluate the conclusions in the paper are present in the paper and/or the Supplementary Materials.\\
\noindent
\textbf{Author contribution:} A.N.M.N.I. and A.S. proposed and conceptualized the project. X.N., J.D., S.K. performed the device fabrication and characterization. The device modeling and network simulation was performed by A.N.M.N.I. All authors contributed to the writing of the manuscript. K.N and A.S. supervised the project.\\

\end{acknowledgement}

\begin{suppinfo}

Supplementary Text: Device Fabrication, Electrical Characterization, Device Operation with Non-zero Floating Gate Voltage, Insights from the Device Model, Abstraction of Dendritic and Somatic Non-linearity, Network Simulation, Exploration of Network Complexity; Polarization switching of the ferroelectric capacitors in the gate stack (Figure S1); Experimental characteristics of the ferroelectric capacitors in the gate stack (Figure S2); Change of Polarization with respect to voltage pulses (Figure S3); Obtained Field Effect Transistor transfer characteristics (Figure S4); Floating gate voltage with respect to applied set voltages to the gates (Figure S5); Impact of lateral scaling (Figure S6); Impact of vertical scaling (Figure S7); Impact of polarization with low activation field (Figure S8); Impact of polarization with high activation field (Figure S9); Abstraction of dendritic non-linearity from floating gate potential (Figure S10); Decision boundary for networks with and without dendrites (Figure S11); Device Model Parameters (Table S1).

\end{suppinfo}

\bibliography{achemso-demo_ARXIV}

\clearpage

\renewcommand{\thefigure}{S\arabic{figure}} 
\setcounter{figure}{0}

\begin{center}

\section{}

\vspace{10mm}
\large Supplementary Information for\\
\large{\textbf{Dendritic Computing with Multi-gate Ferroelectric Field-Effect Transistors}}\\
\vspace{10mm}
\noindent
\normalsize A N M Nafiul Islam, Xuezhong Niu, Jiahui Duan, Shubham Kumar, Kai Ni, Abhronil Sengupta*\\

\vspace{10mm}
\noindent
*Corresponding author(s). Email: sengupta@psu.edu\\
\vspace{10mm}
\noindent
This file includes:\\
\vspace{5mm}

Supplementary Text

Supplementary Figs. S1 to S11

Supplementary Table S1

\end{center}

\clearpage
\section{Supplementary Text}
\subsection{Device Fabrication}

The multi-gate FeFET device is built on a p-type silicon substrate. The n-type source and drain regions are created with ion implantation. A $10nm$ HfO\textsubscript{2} layer deposited with atomic layer deposition (ALD) functions as the gate dielectric. The ferroelectric capacitors incorporate a metal-ferroelectric-metal (MFM) architecture with a W/Hf\textsubscript{0.5}Zr\textsubscript{0.5}O\textsubscript{2}/W stack. First, $50nm$ thick Tungsten bottom electrode was deposited via sputtering. $10nm$ thick ferroelectric Hf\textsubscript{0.5}Zr\textsubscript{0.5}O\textsubscript{2} (HZO) thin films were deposited via ALD at 250°C, utilizing Hf[N(C\textsubscript{2}H\textsubscript{5})CH\textsubscript{3}]\textsubscript{4} and Zr[N(C\textsubscript{2}H\textsubscript{5})CH\textsubscript{3}]\textsubscript{4} as Hafnium and Zirconium precursors, respectively, with oxygen serving as the oxidant. Zirconium concentration was regulated through adjustment of the hafnium-to-zirconium precursor cycle ratio. Following lithographic patterning, $50nm$ thick Tungsten top electrodes were deposited via sputtering. Subsequently, bottom electrode vias were etched to facilitate electrical connections. The orthorhombic ferroelectric phase within the HZO layer was induced through Rapid Thermal Annealing (RTA) at approximately 500°C for 1 minute under Nitrogen atmosphere.

\subsection{Electrical Characterization}

The device characterization is performed with Keithley 4200-SCS Semiconductor Characterization System. All gate capacitors maintained uniform dimensions of $50\mu m \times 50 \mu m$. Multiple polarization states in the ferroelectric gates were achieved through the application of incremental voltage pulses, yielding a broad spectrum of current responses, as illustrated in Figure 2. For parallel programming operations, discrete write pulses were administered independently to individual capacitors connected in parallel. All the capacitors underwent initialization with a $-3V$ reset pulse of $1ms$ duration. Subsequently, pulses with amplitudes ranging from $0V$ to $4V$, in $0.5V$ increments and with consistent $1ms$ duration, were applied. Transfer characteristics were then obtained by varying the gate-source voltage across different polarization states.

\textcolor{black}{\subsection{Device Operation with Non-zero Floating Gate Voltage}}
\textcolor{black}{The case with non-zero charge on the floating gate ($Q_F\neq0$), i.e., $Q_F=\delta,  \delta\neq0$ can be easily derived. We can rewrite Eq. (2) as:}

\textcolor{black}{
\begin{equation}
 C_0\phi_F+\delta=\sum_{i=1}^N(P_i+C_i\cdot V_{FE,i})
\end{equation}}

\textcolor{black}{Solving this gives us:}
\textcolor{black}{
\begin{equation}
\phi_F = \frac{\sum_{i=1}^N C_i \cdot V_i}{\sum_{i=0}^N C_i} + \frac{\sum_{i=1}^N P_i}{\sum_{i=0}^N C_i} - \frac{\delta}{\sum_{i=0}^N C_i}
\end{equation}}

\textcolor{black}{Thus, we get another term in Eq. (3) which is dependent on the residual charge of the floating gate, which impacts the overall floating gate voltage. This will inevitably influence the polarization switching behavior of the ferroelectric gates and thus can be another way to control the device behavior. For the purposes of this work, we have opted to ignore this phenomenon. Likewise, any defects or traps that may impact the charge accumulation and flow in the device (e.g. interfacial traps) will require careful consideration. However, it should be noted that owing to the MFMIS structure adopted by the device, the metal floating gate screens the random spatial distribution of polarization switching. This enables the channel to see an average effect and avoid percolation-related non-idealities \cite{lee2022effect}.}\\

\subsection{Insights from the Device Model}

The calibrated device model allows us to vary various device and material parameters and observe the results to draw greater insights into the device behavior. For example, scaling of the ferroelectric capacitors laterally (i.e., reducing its area) reduces the number of domains that can be incorporated \cite{islam2023hybrid}. Our simulations reveal that although this does make the device’s behavior prone to more variability, the general switching characteristics remain quite similar, so long as the area ratio between the ferroelectric capacitors and FET remains the same (Supplementary Fig. \ref{S6}). This is very encouraging from a scaling point of view and can be easily understood by looking at Eq.3. Reducing the ferroelectric layer thickness (i.e., vertical scaling), however, results in a greater change in behavior (Supplementary Fig. \ref{S7}) under the same operating conditions. Again, referring to Eq. 3, we see that this is owed to a two-fold impact: (i) increase in the effective electric field across the ferroelectric layer, which induces more polarization switching, and (ii) increase of capacitance; both being inversely proportional to thickness. To highlight the effect of only the polarization term, we can analyze different intrinsic activation fields. Supplementary Fig. \ref{S8} and \ref{S9} show the cases where the means of the gaussian activation field distribution are $1MV/cm$ and $5MV/cm$ respectively. Here, a higher nominal value of activation field results in fewer polarization switching events. This yields a smaller polarization term, which in turn makes the device behavior more capacitive in nature, especially when the applied voltage is small (see similarities to Fig. \ref{impact}(d)). In contrast, a lower activation field allows for greater polarization switching even at smaller voltages. 

\textcolor{black}{\subsection{Abstraction of Dendritic and Somatic Non-linearity}}
\textcolor{black}{To extract the dendritic and somatic non-linearity from our device model, we look at the floating gate voltage with respect to one of the gate voltages while setting the others to ground (Supplementary Fig. \ref{S10}). Taking inspiration from the Preisach model of ferroelectric switching \cite{ni2018circuit}, we abstract the branch non-linear function as a hyperbolic tangent. Since only positive input voltages are considered, we set the function equal to zero for all negative values.}

\textcolor{black}{Likewise, the final drain current is another non-linear function of the floating gate potential and acts as the final output of the entire neuron for that column. For this, a rectified linear unit (ReLU) is selected (Supplementary Fig. \ref{S4}) for training our neural network.} \textcolor{black}{It should be noted that these abstractions, both for the dendritic and somatic non-linearity, is done to efficiently train the network. The inference afterwards is performed directly with the device dynamics that considers complexities like concurrent non-zero voltages applied to multiple gates, among others.} \\

\subsection{Network Simulation}

The network is constructed in Python using the PyTorch library \cite{paszke2019pytorch}. We train the network using the Adam optimizer \cite{kingma2017adammethodstochasticoptimization} and employ the cross-entropy function to calculate the loss. Batch normalization \cite{ioffe2015batch} and dropout of 0.5 \cite{srivastava2014dropout} were utilized for regularization. Both versions of the network (with and without dendrites) were trained over 100 epochs with a learning rate that is exponentially decaying from $10^{-2}$ to $10^{-5}$. We used a batch-size of 64 for the training process.

\textcolor{black}{During inference, as we are interested in the device performance, we assumed ideal software crossbars and mapped the crossbar outputs directly to voltages, which were then input to each of the gates of our device. This mapping ensures we can utilize the entire switching range (linear mapping, an output of zero from the crossbar maps to 0V and the maximum obtained software output maps to $4V$).} \textcolor{black}{Additionally, a $20\%$ variation in output was assumed for the different neurons of the network to take into account device-to-device variability. While variability at a per capacitor level was not included in our analysis, this general catch-all term is used to model variations in the system. Further investigation of variation based on extensive experimentation and more granular considerations are left for future exploration. }

\textcolor{black}{\subsection{Exploration of Network Complexity}}
\textcolor{black}{To explain the apparent advantage owing to using dendritic networks, we explore how dendrites help to increase the network’s complexity or expressivity. It has been shown that non-linear perceptrons divide the input space by hyperplanes \cite{montufar2014number}. Generally, the more hyperplanes dividing the input space, the greater the network's complexity \cite{raghu2017expressive}. To explore this empirically using visuals, we train a small network on scikit-learn's \cite{pedregosa2011scikit} concentric circles dataset, where data points are sampled from two circles of differing radius. It is a classic 2D non-linear problem, which cannot be effectively solved with a linear network. We train two versions of small networks with 2 input neurons, 2 hidden layer neurons and 2 output neurons, one with dendrites ($k=2$) and the other without ($k=0$). After training, we plot the decision boundary of the two networks along with the test dataset, as shown in Supplementary Fig. \ref{S11}. We find that the network without dendrites is only able to separate the input space with 2 planes, whereas the network with dendrites is able to more accurately wrap around the inner circle to differentiate the two sets of points. Although this is a very simplistic simulation, it still elucidates the increased complexity afforded by the non-linearity of dendrites that lies at the center of this improvement, i.e. such a functionality would not be possible without the ferroelectric layer.}

\clearpage
\begin{figure}
\centering
\includegraphics[width=\textwidth]{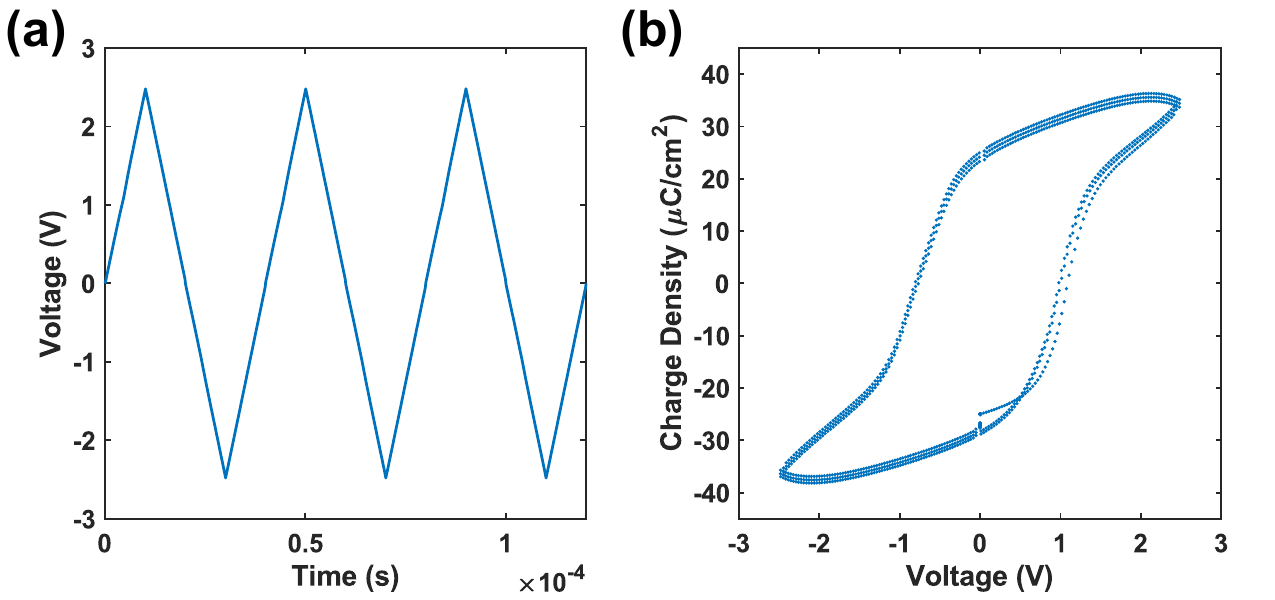}
\caption{\textcolor{black}{Polarization switching observed for the ferroelectric capacitors in the gate stack. (a) Input voltage to the capacitor is a triangular wave with an amplitude of $2.5V$ and time period of $40\mu s$. (b) Charge vs voltage characteristics of the ferroelectric capacitor. The hysteresis indicates the ferroelectric polarization switching.}}\label{S1}
\end{figure}

\clearpage
\begin{figure}
\centering
\includegraphics[width=\textwidth]{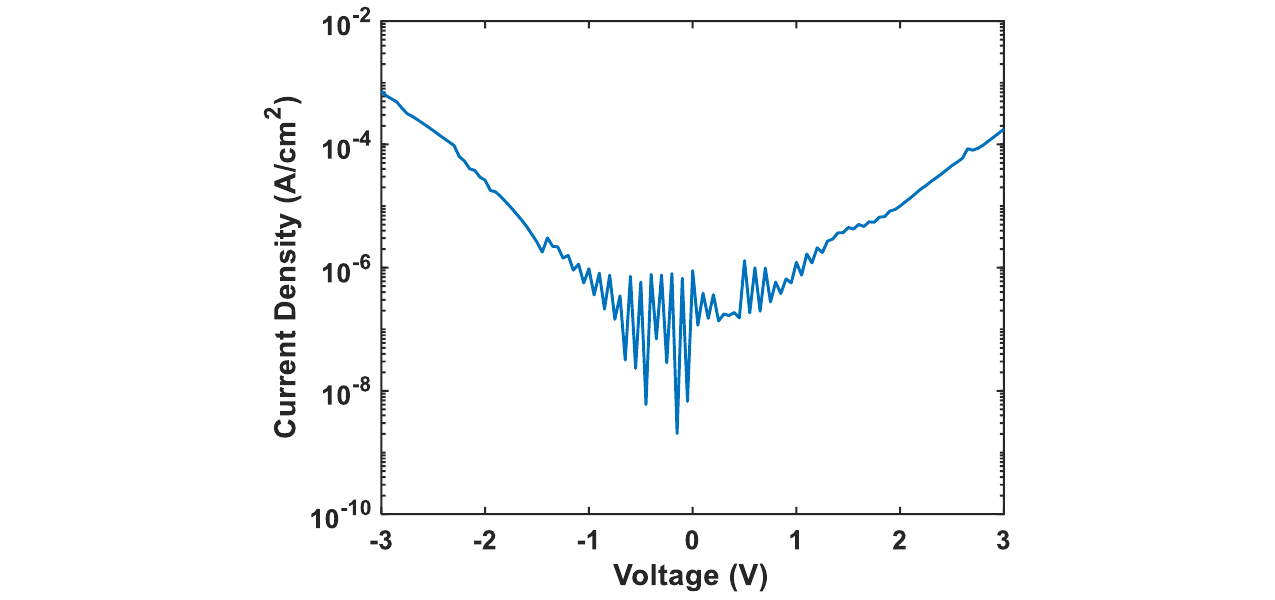}
\caption{\textcolor{black}{Experimental DC current-voltage characteristics of the ferroelectric capacitors in the gate stack.}}\label{S2}
\end{figure}

\clearpage
\begin{figure}
\centering
\includegraphics[width=\textwidth]{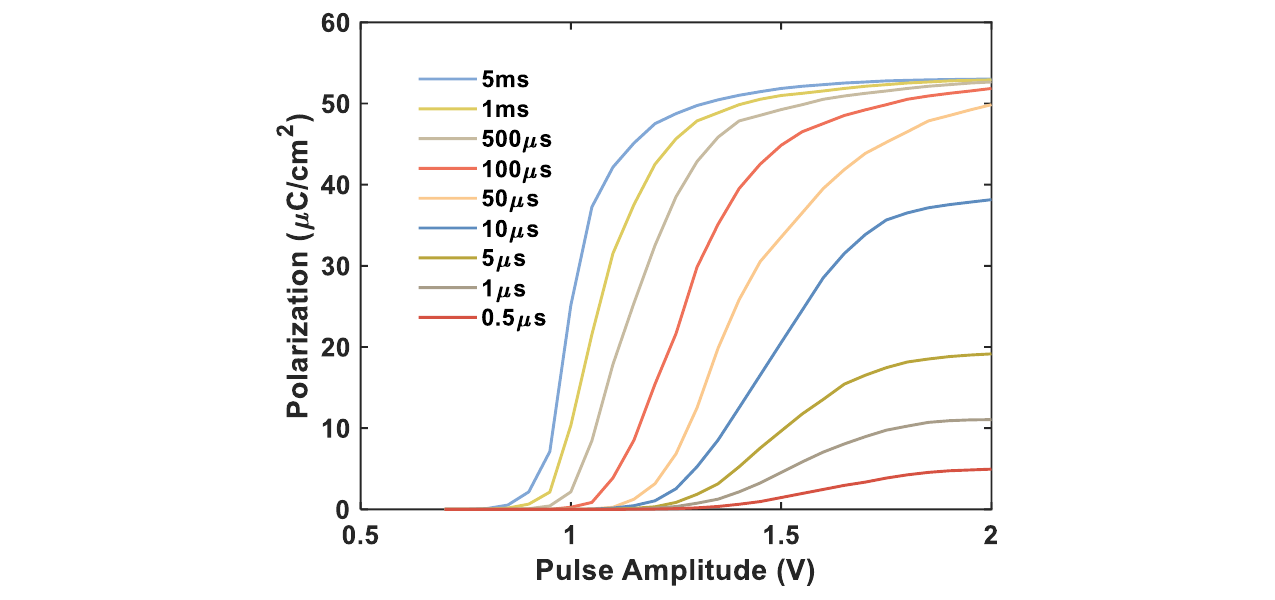}
\caption{\textcolor{black}{Change of Polarization with respect to voltage pulses of different amplitudes and widths.}}\label{S3}
\end{figure}

\clearpage
\begin{figure}
\centering
\includegraphics[width=\textwidth]{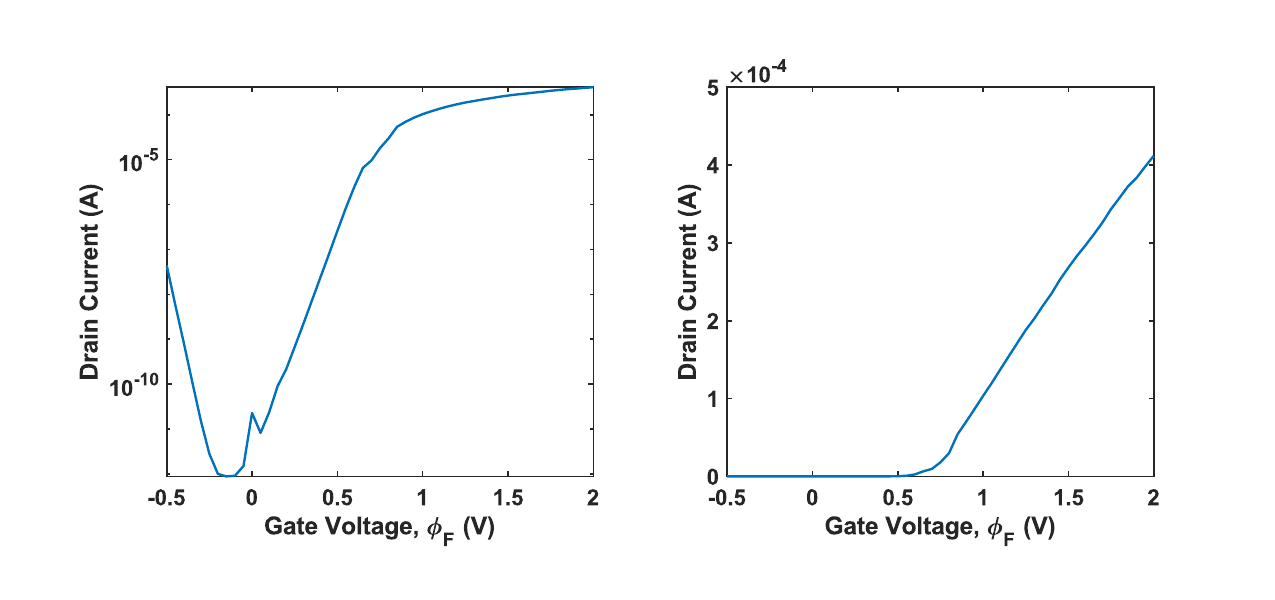}
\caption{Field Effect Transistor transfer characteristics obtained by sweeping the gate voltage $\phi_F$ from $-0.5V$ to $2V$ and measuring the drain current with an applied drain voltage of $0.1V$ in logarithmic scale (left) and linear scale (right).}\label{S4}
\end{figure}

\clearpage
\begin{figure}
\centering
\includegraphics[width=\textwidth]{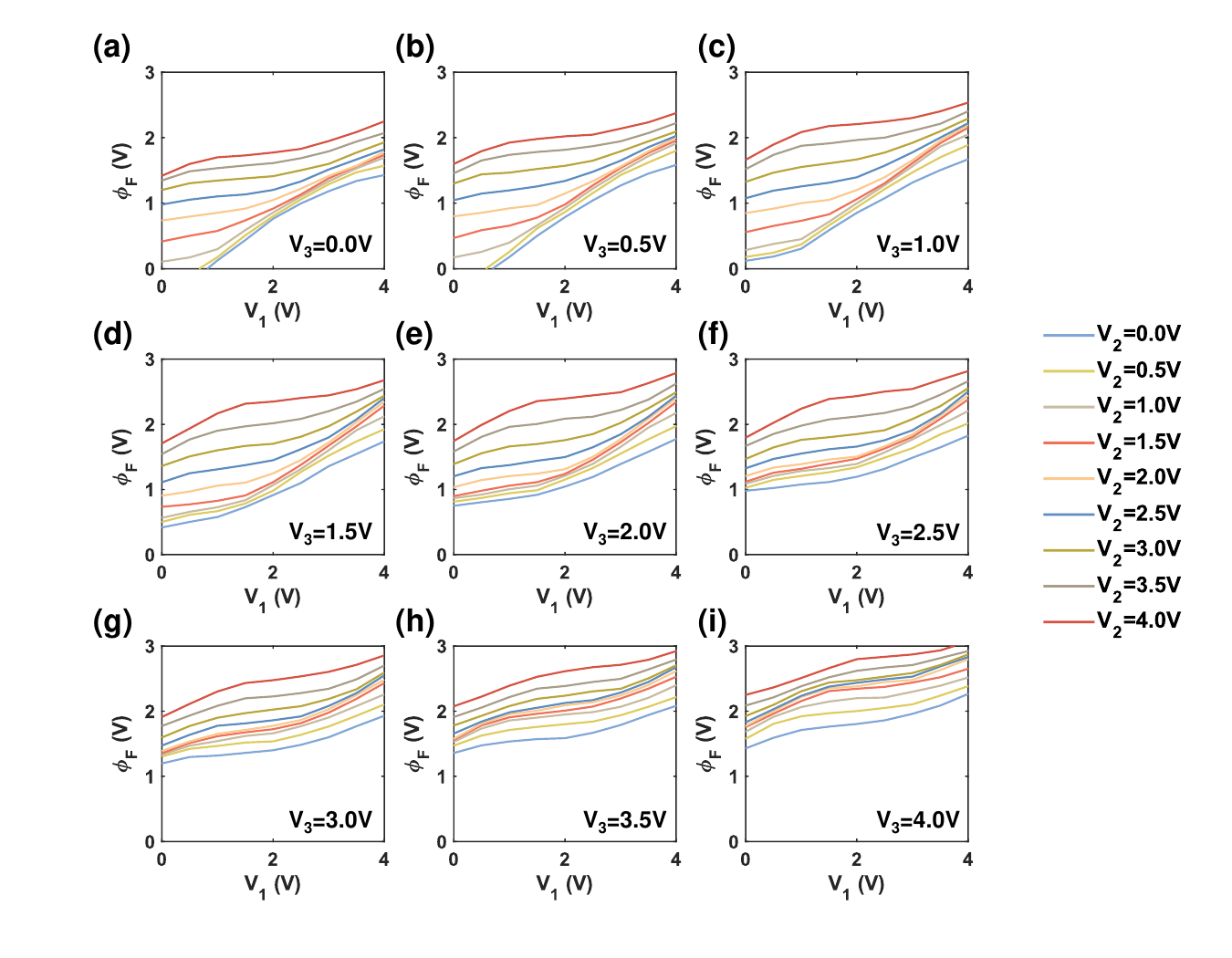}
\caption{Floating gate voltage, $\phi_F$ with respect to applied set voltage to the first gate, $V_1$ for the multi-gate FeFET corresponding to same conditions as Fig. 2(c-k) of the main text. (a-i) For each panel, the different colored curves represent different set pulse values applied to the second gate, $V_2$. Likewise, the set voltage applied to the third terminal, $V_3$ remains constant for each panel and is noted at the bottom right of the panel. }\label{S5}
\end{figure}

\clearpage
\begin{figure}
\centering
\includegraphics[width=\textwidth]{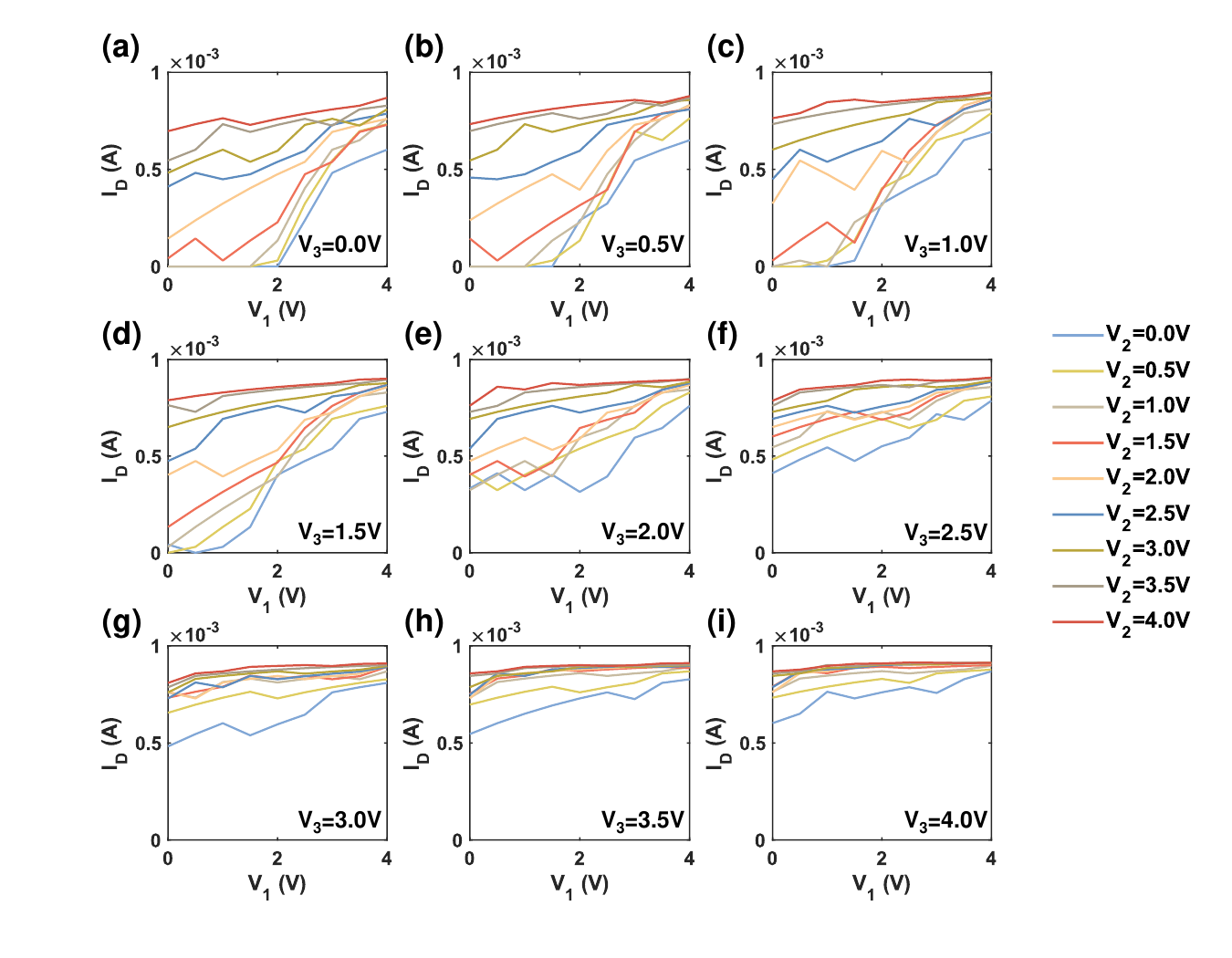}
\caption{Drain current $I_D$ with respect to applied set voltage to the first gate, $V_1$ for a multi-gate FeFET with scaled ferroelectric capacitors. For the simulation, we chose the number of domains, $N_{dom}$ to be 20, according to findings in \cite{islam2023hybrid}. (a-i) For each panel, the different colored curves represent different set pulse values applied to the second gate, $V_2$. Likewise, the set voltage applied to the third terminal, $V_3$ remains constant for each panel and is noted at the bottom right of the panel. We find that although the device behavior becomes somewhat erratic, the trend in the characteristics is preserved even when we consider scaling.}\label{S6}
\end{figure}

\clearpage
\begin{figure}
\centering
\includegraphics[width=\textwidth]{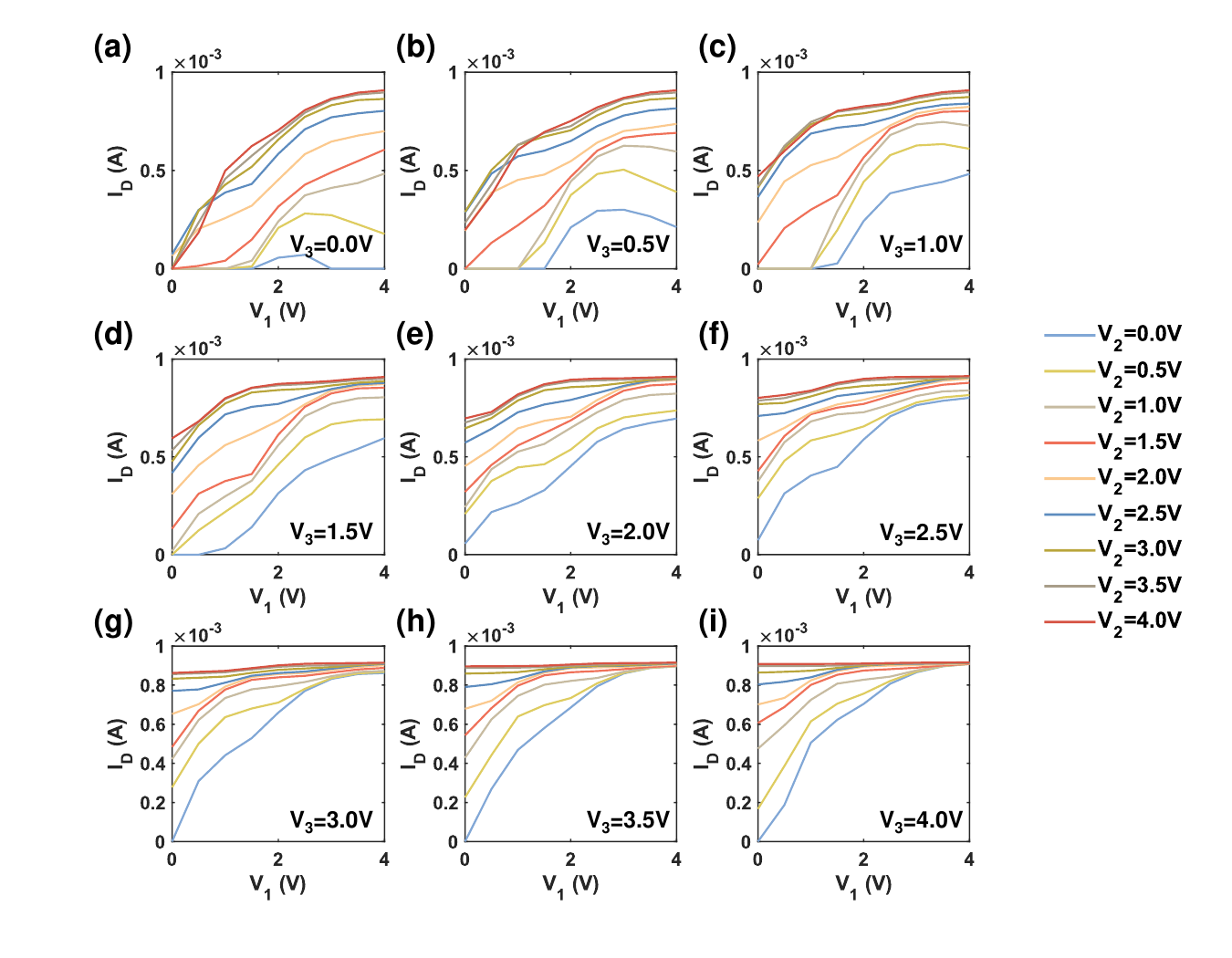}
\caption{Drain current $I_D$ with respect to applied set voltage to the first gate, $V_1$ for a multi-gate FeFET with scaled ferroelectric capacitors. For the simulation, we chose the ferroelectric layer thickness, $t_{FE}$ to be $5nm$. (a-i) For each panel, the different colored curves represent different set pulse values applied to the second gate, $V_2$. Likewise, the set voltage applied to the third terminal, $V_3$ remains constant for each panel and is noted at the bottom right of the panel. }\label{S7}
\end{figure}

\clearpage
\begin{figure}
\centering
\includegraphics[width=\textwidth]{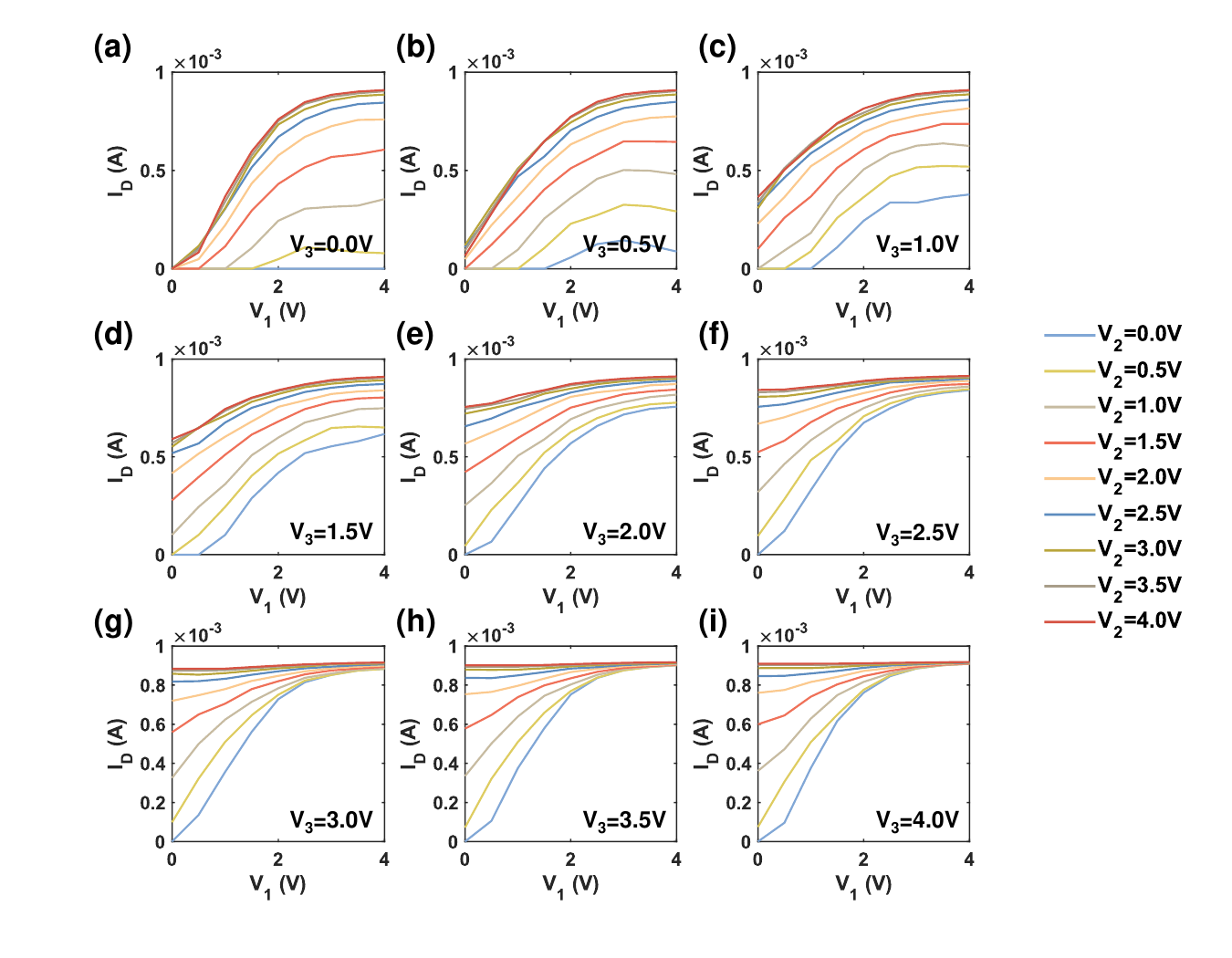}
\caption{Drain current $I_D$ with respect to applied set voltage to the first gate, $V_1$ for a multi-gate FeFET with scaled ferroelectric capacitors. For the simulation, we chose the mean of the activation field to be $1MV/cm$. (a-i) For each panel, the different colored curves represent different set pulse values applied to the second gate, $V_2$. Likewise, the set voltage applied to the third terminal, $V_3$ remains constant for each panel and is noted at the bottom right of the panel.}\label{S8}
\end{figure}

\clearpage
\begin{figure}
\centering
\includegraphics[width=\textwidth]{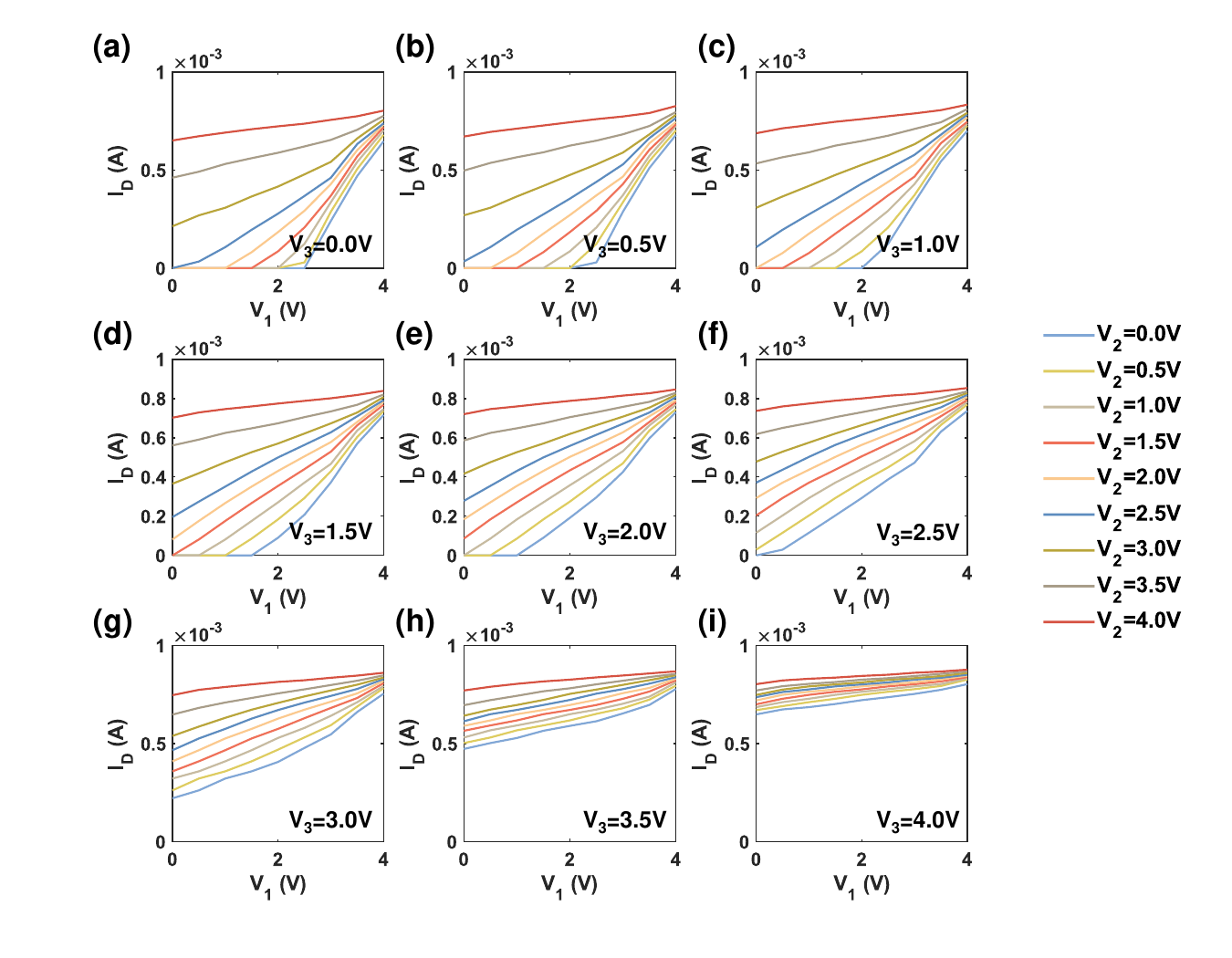}
\caption{Drain current $I_D$ with respect to applied set voltage to the first gate, $V_1$ for a multi-gate FeFET with scaled ferroelectric capacitors. For the simulation, we chose the mean of the activation field to be $5MV/cm$. (a-i) For each panel, the different colored curves represent different set pulse values applied to the second gate, $V_2$. Likewise, the set voltage applied to the third terminal, $V_3$ remains constant for each panel and is noted at the bottom right of the panel.}\label{S9}
\end{figure}

\clearpage
\begin{figure}
\centering
\includegraphics[width=\textwidth]{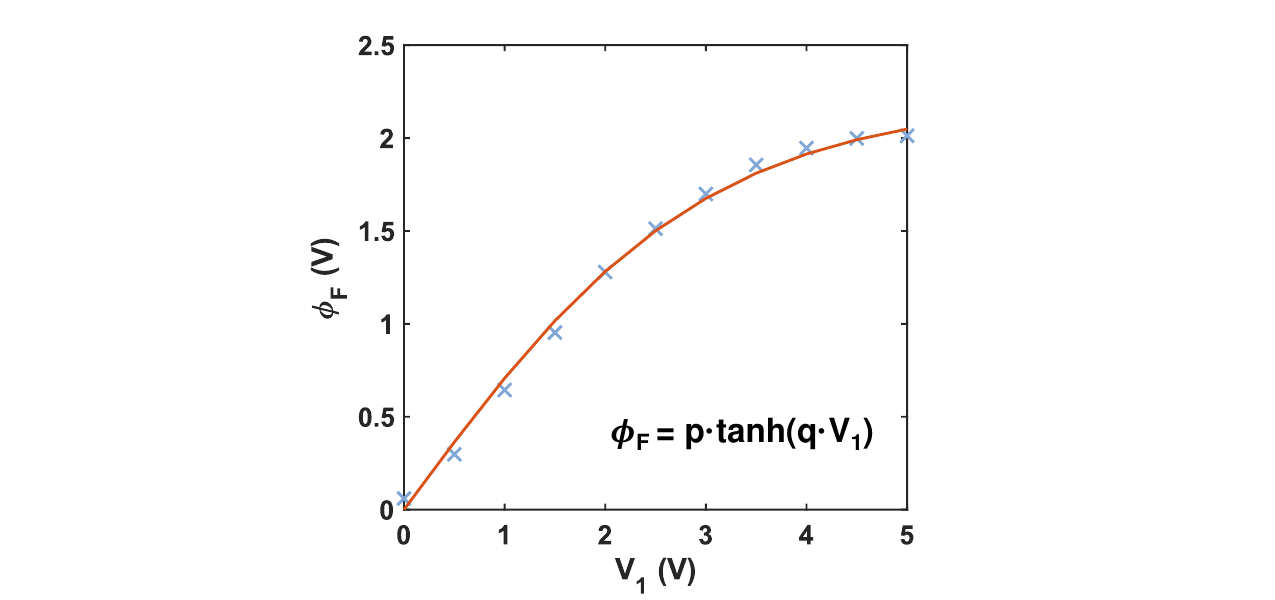}
\caption{With the calibrated model, we extract the floating gate voltage, $\phi_F$ with respect to one of the gate voltages, $V_1$, when $V_2/V_3=0$. Taking inspiration from the Preisach model of ferroelectric switching \cite{ni2018circuit}, we fit the data to the equation of form, $\phi_F=p \cdot tanh(q \cdot V_1)$, where $p=2.2$ and $q=1/3$. We use this abstraction as the branch non-linearity to train our network.}\label{S10}
\end{figure}

\clearpage
\begin{figure}
\centering
\includegraphics[width=\textwidth]{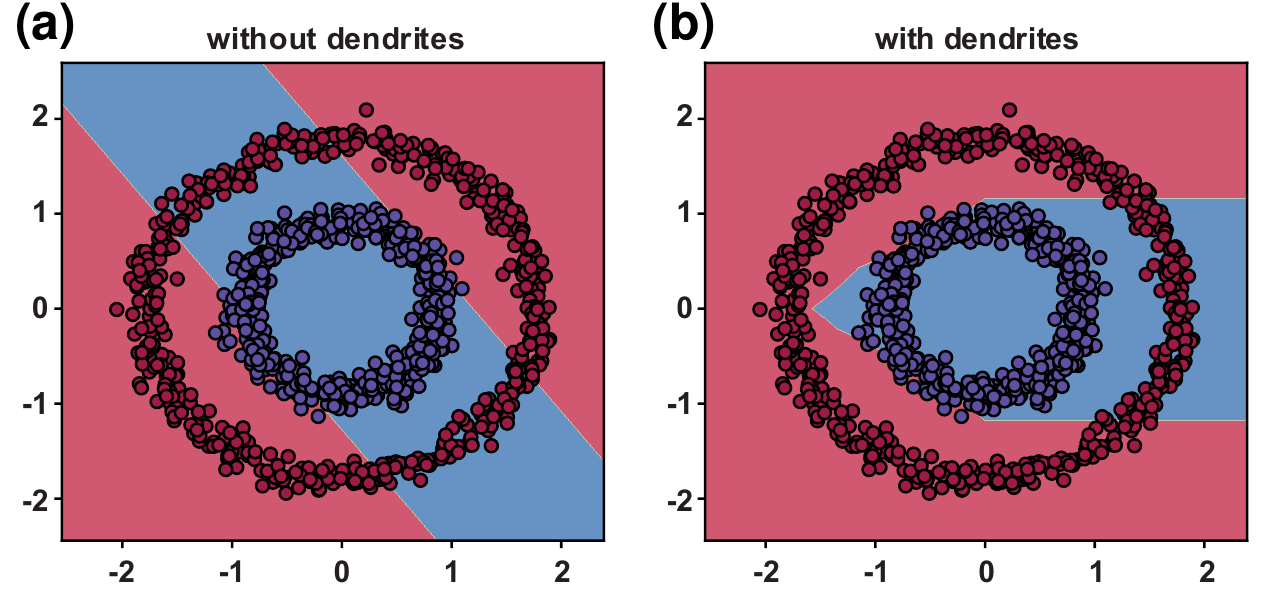}
\caption{Decision boundary for the concentric circles dataset with two classes for (a) a network without dendrites, $k = 0$  and (b) a network with dendrites, $k = 2$.}\label{S11}
\end{figure}

\clearpage

\renewcommand{\thetable}{S\arabic{table}} 
\begin{table}[h]
\caption{Device Model Parameters}
\vspace{3ex}
    \centering  
    \begin{tabular}{c c}
    \hline
    \textbf{Parameters} & \textbf{Value}\\
    \hline
    
     Number of domains, $N_{dom}$ & $1000$\\
     Time-step, $\Delta t$ &  $10\mu s$\\
     Saturated polarization, $P_S$ & \ $28\mu C/cm^2$  \\
     Polarization time constant, $\tau_0$, $\alpha$ &  $2\times10^{-8}s$, $4.0$\\
     Activation field distribution, $E_a$ &  $\mathcal{N}(\mu=3.0,\sigma=1.0)$\\
     Shape parameter, $\beta$ &  $2$\\
     Temperature &  $300K$\\
     Substrate Doping &  $1\times 10^{15}cm^{-3}$\\
     FET oxide thickness &  $10nm$ \\
     Ferroelectric capacitor thickness, $t_{FE}$ &  $10nm$ \\
     Read drain voltage &  $0.1V$\\
     Read source voltage &  $0V$\\
        
    \hline
    \end{tabular}    
    \label{dev_params}
\end{table}

\clearpage


\end{document}